\documentclass[usenatbib]{mn2e}

\usepackage[dvipdf,dvips]{graphicx}
\usepackage{amsmath}

\def\tcdm{$\tau$CDM }
\def\lcdm{$\Lambda$CDM }

\bibpunct[,]{(}{)}{;}{a}{}{,}

\begin{document} 

\title[Higher-Order Moments of Dark Halo Distribution]
{Testing Theoretical Models for the Higher-Order Moments
of Dark Halo Distribution}
\author[R. Casas-Miranda \textit{et\thinspace al\/}]
{R. Casas-Miranda, H.J. Mo and G. Boerner\\
Max-Planck-Institut f\"ur Astrophysik, Garching, Germany
}
\date{}
\maketitle

\begin{abstract}
Using high--resolution N--body simulations, we test two theoretical
models, based either on spherical or on ellipsoidal collapse model,
for the higher--order moments of the dark matter halo distribution in
CDM models.
We find that a theoretical model based on spherical collapse describes
accurately the simulated counts--in--cells moments for haloes of
several mass ranges. It appears that the model using ellipsoidal
collapse instead of spherical collapse in defining dark haloes is
unable to improve the models for the higher--order moments of halo
distribution, for  haloes much smaller than $M^*$ (the mass scale on which the fluctuation
of the density field has a rms about 1).
Both models are particularly accurate for the descendants of haloes 
selected at high redshift, and so are quite useful in 
interpreting the high--order moments of galaxies. 
As an application we use the theoretical
model to predict the higher--order moments of the Lyman break 
galaxies observed at $z\approx 3$ and their descendants 
at lower redshifts. 
\end{abstract}


\section{Introduction}
In the standard scenario of galaxy formation, it is assumed that 
galaxies form by the cooling and condensation of gas within dark
haloes \citep[e.g.][]{WhiteR:78, WhiteF:91}. The problem of
galaxy clustering in space can then be approached by
understanding the spatial distribution of dark haloes 
and galaxy formation in individual dark haloes. This
approach is very useful for the following two reasons:
(i) the formation and clustering properties of dark haloes 
can be modelled relatively reliably because of the simple physics 
involved (gravity only), (ii) realistic models of 
galaxy formation in dark haloes can now be constructed using 
either semi-analytic models \citep[e.g.][]{Kauffmann:99, Cole:00,
SomervilleP:99} or hydrodynamical simulations \citep[e.g.][]{Benson:01}. 
Indeed, attempts have been made to use theoretical models
of halo clustering to understand clustering properties of galaxies
\citep[e.g.][]{MoJB:97, JingMB:98, MaF:00, Scoccimarro:01,
PeacockS:00, Seljak:00}. 
Most of these investigations use the theoretical models
presented in \citet[][]{MoW:96} and in \citet[][]{MoJW:97} (hereafter
MJW) to calculate the second-order and higher-order correlations of  
dark haloes. These models are based on the Press-Schechter 
formalism \citep[][]{PressS:74} and its extensions \citep[][]{LaceyC:94}.

 The model prediction for the second moment, or the two-point
correlation function, has been tested quite extensively  
by numerical simulations \citep[][]{MoW:96, MoJW:96, Jing:98,
ShethT:99, Governato:99, Colberg:00}.
The results show that the model proposed by Mo \& White 
works reasonably well over a large range of halo masses.
However, significant discrepancy between model and 
simulations results was found for low-mass haloes \citep[][]{Jing:98,
ShethT:99}. \citet[][]{ShethMT:01} (hereafter SMT) suggested that the
discrepancy 
at the low-mass end may be due to the fact that the model 
considered by Mo and White assumes spherical collapse
for the halo formation while the collapse in a realistic
cosmological density field may be better approximated 
by an ellipsoidal model. Indeed, SMT found that,
if an ellipsoidal model is used, better agreement between 
the model and simulations results can be achieved 
in both the halo mass function and the two-point 
correlation function for low-mass haloes.

The performance of the MJW model for the higher-order 
moments of the halo distribution has been tested in their 
original paper using scale-free N-body simulations 
with relatively low resolution. Although their results
show that the theoretical model matches the simulations 
results, the limited dynamical range in the simulations
used by them does not allow one to test the model for
a large range of halo masses. Furthermore, although the 
MJW model has been extended to include ellipsoidal 
dynamics \citep[][]{ShethMT:01}, this extension has
not yet been tested by simulations results. 
     
 In this paper we use two sets of high-resolution simulations 
to test the MJW model and its extension. One set has a 
very large simulation box (and thus low mass resolution)
which is used to control the finite-volume effect usually
found in the analysis of higher-order moments of the galaxy distribution
\citep[][]{ColombiBS:94}. The other set has smaller simulation
boxes but much higher mass resolutions which allows us to 
test the models for low-mass haloes. 

The paper is organized as follows: The procedure to obtain the high
order moments from counts-in-Cells is presented in section 2
along with theoretical models of these moments for
dark haloes. Analysis of the simulations data and the comparison 
of the theory with the simulations results are presented in
section 3. An application of the theoretical models to the
higher-order moments of the Lyman-break
galaxies (LBGs) is done in section 4. Because LBGs are highly
biased \citep[][]{Adelberger:98}, the skewness and kurtosis
coefficients ($S_3$ and $s_4$) of LBGs and their descendants are
significantly lower than those of the mass.

Finally, section 5 contains
a summary of our results.

\section{Higher-Order Moments of Counts-in-Cells}

\subsection{Definitions}
\label{cic_moments}

The calculation of the counts-in-cells
moments of a particle distribution and the relation
of such moments to the corresponding moments of the underlying 
continuous density field are described in detail 
in \citet[][]{Peebles:80}. We summarize the relevant formulae 
in the following.

The $j^{th}$ central moment of counts in cells 
(we will use spherical cells whose radius will be 
denoted by $R$) of a point distribution is defined as 
\begin{equation}
m _j (R)= \frac{1}{M}\sum_{i=1} ^M {(N_i - \bar{N})^j },
\end{equation}
where $N_i$ is the number of particles counted in the $i^{th}$ sphere
(cell), $\bar{N}$ is the mean number of counts:
$\bar{N} (R)= \frac {1}{M} \sum _{i=1} ^{M} {N_i (R)}$, and the
summation is over the $M$ sampling spheres. Notice that
$\bar{N}$ is obtained directly from the counts.

The connected moments, $\mu _i$, 
are defined through the central moments as
\begin{eqnarray}
\mu _2 &=& m_2\,, \\
\mu _3 &=& m_3\,, \\
\mu _4 &=& m_4 - 3 m_2 ^2\,.
\end{eqnarray}
These relations are written up to the 4th order because to this order
they are relevant to our subsequent discussion.
For a point process, the shot noise also contributes to the
quantities $\mu _j$. These contributions become significant 
for small radius where the mean count $\bar{N}$ is small
and should be properly subtracted. If the particle distribution
is a Poisson sampling of the underlying density distribution,
we can make the following subtractions to get the corrected
connected moments:
\begin{equation}
k_2 = \mu_2 - \bar{N}, 
\end{equation}
\begin{equation}
k_3 = \mu_3 - 3 \mu_2 + 2 \bar{N},
\end{equation}
\begin{equation}
k_4 = \mu_4 - 6 \mu_3 + 11 \mu_2 - 6 \bar{N}.
\end{equation}
These quantities are related to the volume-averaged correlation 
functions by 
\begin{equation}
\bar{N}^j \bar{\xi}_j = k_j\,,
\end{equation}
where 
\begin{equation}
\bar{\xi}_j =
V_W^{-j} \int  dr_1...dr_j ~W(r_1)...W(r_j)~ \xi_j(r_1,...,r_j)\,,
\end{equation}
and $W(r)$ is a top-hat spherical window with volume $V_W$. 

\subsection{Theoretical Model for the Higher-Order Moments of Dark Halos}

\citet[][]{MoJW:97} (MJW) have developed an analytical
model for the hierarchical correlation amplitudes [$S_{j,h}(R) =
\bar{\xi}_{j,h}/\bar{\xi}_{2,h}^{j-1}$] for $j=3,4,5$ in the 
quasi-linear regime, where the subscript $h$ stands for quantities 
of dark haloes. In this model the statistical distribution of 
dark haloes within the initial density field, which is assumed 
to be Gaussian, is determined
by an extension of the Press-Schechter formalism and the modifications
of the distribution due to gravitationally induced motions are treated
by means of a spherical collapse model \citep[][]{MoW:96}. 
The main results from this
model, which are relevant for our analysis, are summarized as
follows.

For the skewness and kurtosis ($j=3~,4$), following the same
notation as in MJW, one has:
\begin{equation}
S_{3,h} = b^{-1}(S_3+3c_2),
\label{s3_h}
\end{equation}
\begin{equation}
S_{4,h} = b^{-2}(S_4 +12c_2S_3 +4c_3+12c_2^2),
\label{s4_h}
\end{equation}
where $S_3$ and $S_4$ are the skewness and kurtosis of the underlying
mass density field, $c_k=b_k/b$,  $b=b_1$, and the constants $b_k$ are the
coefficients in the expansion of the bias relation: 
\begin{equation}
\delta_h = \sum_{k=0}^\infty {\frac{b_k}{k!}}\delta^k ,
\end{equation}
where $\delta_h$ is the overdensity of haloes smoothed in a given
window and $\delta$ is the corresponding overdensity of mass.
The coefficients $b_k$ for a halo with mass $M_1$ corresponding to a
linear overdensity $\delta _1$, which collapses at redshift $z_1 =
\delta _1/\delta _c -1$ (with the critical overdensity for spherical collapse
being $\delta _c = 1.686$), are given by:
\begin{equation}
b_1 = 1+{\frac{\nu_1^2-1}{\delta_1}},
\label{b1}
\end{equation}
\begin{equation}
b_2 = 2(1+a_2){\frac{\nu_1^2-1}{\delta_1}}+
\left(\frac{\nu_1}{\delta_1}\right)^2(\nu_1^2-3),
\label{b2}
\end{equation}
\begin{eqnarray}
b_3 &=& 6(a_2+a_3){\frac{\nu_1^2-1}{\delta_1}} 
   +3(1+2a_2)\left({\frac{\nu_1}{\delta_1}}\right)^2(\nu_1^2-3) \nonumber\\
   &+& \left({\frac{\nu_1}{\delta_1}}\right)^2 
       {\frac{\nu_1^4-6\nu_1^2+3}{\delta_1}},
\label{b3}
\end{eqnarray}
where $\nu_1\equiv \delta_1/\sigma(M_1)$ 
[with $\sigma(M_1)$ being the rms of the density fluctuation
given by the density spectrum linearly extrapolated to the present 
time], $a_2 = -{\frac{17}{21}} $ and $a_3 = {\frac{341}{567}}$
are coefficients in the expansion of $\delta_0(\delta)$,
the relation between the real mass overdensity $\delta$ and 
the corresponding quantity obtained using linear theory  
(see equation A4 in MJW). The above bias coefficients
are for the present-day descendants (at redshift $z_0=0$) 
of haloes identified at redshift $z_1$. It can be easily extended to 
the case where $z_1>z_0>0$. In this case, we replace
$\delta_1$ by $\delta_1 D(z_0)/D(0)$ (where $D(z)$ is the 
linear growth rate evaluated at redshift $z$) while keeping 
$\nu_1$ unchanged. 

If an ellipsoidal model is used to define collapsed haloes,
the coefficients $b_k$ take the following forms
\citep[][]{Scoccimarro:01}: 
\begin{equation}
b_1 = 1 + \epsilon_1 + E_1,
\label{b1_st}
\end{equation}
\begin{equation}
b_2 = 2(1+a_2) (\epsilon_1 + E_1) + \epsilon_2 + E_2,
\label{b2_st}
\end{equation} 
\begin{equation}
b_3 = 6(a_2 + a_3)\,(\epsilon_1+E_1) 
         + 3(1 + 2a_2)\,(\epsilon_2 + E_2) + \epsilon_3 + E_3,
\label{b3_st}
\end{equation} 
where 
\begin{eqnarray}
\epsilon_1 &=& \frac{\alpha \nu^2-1}{\delta_1}, \nonumber\\ 
\epsilon_2 &=& \frac{\alpha\nu^2}{\delta_1^2} (\alpha \nu^2-3),\\
\epsilon_3 &=& \frac{\alpha\nu^2}{\delta_1^3} (\alpha^2\nu^4 - 6\alpha\nu^2
+ 3),\nonumber\\ 
\end{eqnarray}
\begin{eqnarray}
E_1 &=&  \frac{2p/\delta_1}{1+ (\alpha \nu^2)^p},\nonumber\\
\frac{E_2}{E_1} &=& \frac{1+2p}{\delta_1} + 2\epsilon_1, \\
\frac{E_3}{E_1} &=& {4(p^2-1) + \frac{6p\alpha\nu^2}{\delta_1^2}} +
3\epsilon_1^2,\\ 
\end{eqnarray}
and $\alpha=0.707$, $p=0.3$. These formulae reduce to
the original MJW model for $\alpha=1$ and $p=0$. 

Inserting the expressions for $b_k$ in equations (\ref{s3_h}) and
(\ref{s4_h}) and taking $S_3$ and $S_4$ in these equations 
as the skewness and kurtosis of the mass distribution 
measured directly from the N-Body simulations, we
can calculate the skewness and kurtosis for the distribution of dark
haloes as predicted by the MJW model and its SMT extension
(i.e. the model with ellipsoidal collapse).

\section{Skewness and Kurtosis of Dark Haloes in N-body Simulations}
        
\subsection{Simulations}
        
In the present analysis we use two sets of cosmological 
N-body simulations, which have been obtained as part of the VIRGO
\citep[][]{Jenkins:98} and the GIF \citep[][]{Kauffmann:99}
projects. These two sets of simulations differ in the size of the simulation
boxes and in the mass resolution, with the VIRGO simulations having larger
simulation boxes and lower mass resolutions than the GIF ones. From the
VIRGO simulations we have analyzed the \lcdm model in order to test
the MJW model in a volume large enough so that the effects due to
the finite sampling volume may be negligible (see below). 
We compare the results with those obtained from the GIF simulations 
to see how comparisons between models and simulations can be made 
for simulations with relatively small volume. 
For the GIF simulations, we focus on the \tcdm and
\lcdm models. The parameters characterizing the simulations 
are summarized in table 1. Further details can be found in 
\citet[][]{Kauffmann:99} and \citet[][]{Jenkins:98}.

\begin{table*}
\label{tab_par}
\centering
\begin{tabular}{l*{9}{c}}
\hline
Model & $\Omega_0$ & $\Omega_\Lambda$ & $h$ & $\sigma_8$ &
$\Gamma$ & Box Size & $N_{\rm p}$ & $m_p/h^{-1} M_{\odot}$ \\
& & & & & & [${\rm Mpc}/h$]  \\
\hline
GIF-$\tau$CDM      & 1.0 & 0.0 & 0.5 & 0.60 & 0.21 &  85  & $256^3$ 
                   &$1.0 \times 10^{10}$\\
GIF-$\Lambda$CDM   & 0.3 & 0.7 & 0.7 & 0.90 & 0.21 & 141  & $256^3$
                   &$1.4 \times 10^{10}$\\
\hline
VIRGO-$\Lambda$CDM & 0.3 & 0.7 & 0.7 & 0.90 & 0.21 & 239.5& $256^3$ 
                   &$6.86\times 10^{10}$\\
\hline
\end{tabular}
\caption{Parameters characterizing the simulations used in the paper.
$\Omega_0$ and $\Omega_\Lambda$ are the density parameters for matter 
  and for the cosmological constant, respectively, 
  $h$ is the Hubble parameter, $\sigma_8$ is
  the rms of the density field fluctuations in spheres of radius
  $8\,h^{-1}\;$Mpc, and $\Gamma$ is the shape parameter of the power
  spectrum. Also given are the size of the simulation box, the
  total number of particles and the mass per dark matter particle 
  in a simulation.}
\end{table*}

For each simulation there are several output files corresponding to
different evolutionary times (redshifts) and for each of these 
output times there is a halo catalog containing information 
about haloes identified using the friends-of-friends group-finder algorithm
with a linking length 0.2 times the mean interparticle separation.
Only haloes containing 10 or more particles are included in the
halo catalogues. In the case of the GIF simulations at redshift
zero, which are used to compare the predictions of the models for 
haloes with low masses, unbound objects are excluded from the 
catalogues (see section 3.2 for details). The physical quantities 
available from each of these halo catalogues are: the index of the most-bound
particle in the halo, which corresponds to the position of the halo as
well as the central `galaxy' within it; the virial radius
$(R_{vir})$, defined as the radius (from the central particle) within
which the overdensity of dark matter is 200 times the critical
density; the virial mass $(M_{vir})$, which is the mass 
(or, equivalently, the total number) of dark matter
particles within the virial radius; the circular velocity [$V_c =
(GM_{vir}/R_{vir})^{1/2}$].  

We have also generated several catalogues of the present-day
positions of the central objects corresponding to the most-bound particles
in haloes identified at an earlier epoch. This catalogues might be
interpreted as `galaxy catalogues' if we assume that the positions
of galaxies at the present epoch correspond to those of the 
central particles within virialized objects identified at high
redshifts. This concept is related to the assumption in models of 
galaxy formation that galaxies form by the cooling and condensation 
of gas within dark matter haloes \citep[][]{WhiteF:91}. However, 
this interpretation does not take into account subsequent galaxy mergers.

We apply the counts-in-cells analysis described in the last section
to the mass distributions and halo catalogues.
To do this, we place spheres in a regular mesh of 
$30^3$ centers and count the number of objects
at each center over a set of concentric spheres which
allows us to compute the desired statistical quantities at different
radii. 

\subsection{Results}

Following the procedure given in section \ref{cic_moments} we have
obtained the volume-averaged correlation functions up to the fourth order
from the mass distribution and from the various halo samples.
Analyses have been performed for two different cases.
In the first, the higher-order moments are calculated 
at the same time when the dark haloes are identified.
In the second, haloes are identified at some high redshift
while the calculations of the higher-order moments are performed
for their descendants at a later time.
In all cases, the redshift at which halo identification
is made is denoted by $z_1$, while the redshift at which 
the higher-order moments are calculated is denoted by $z_0$.
 
\begin{figure*}
\centerline{
\includegraphics*[bb= 35 160 570 700, width=0.51\textwidth]
{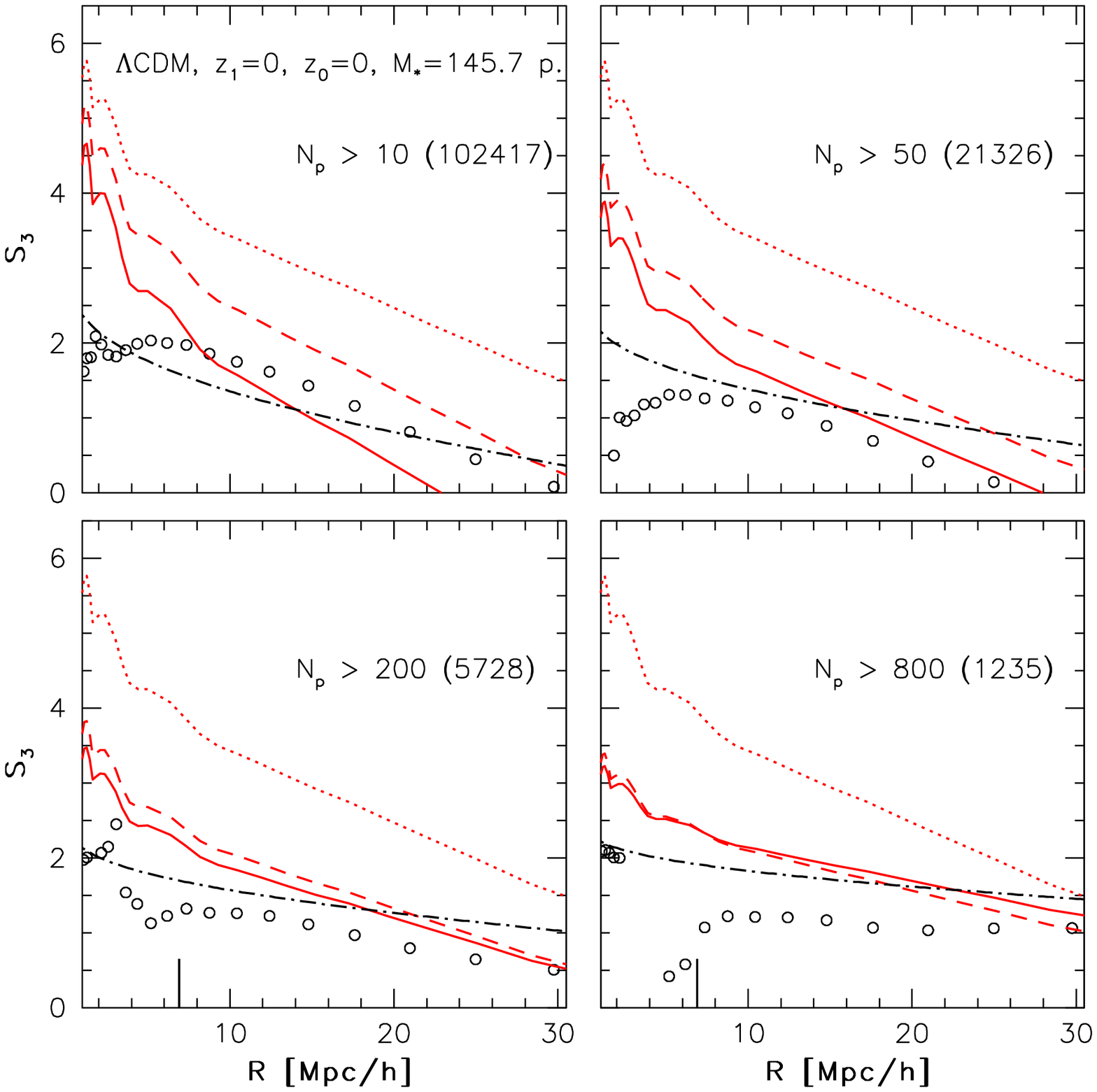}
\includegraphics*[bb= 35 160 570 700, width=0.51\textwidth]
{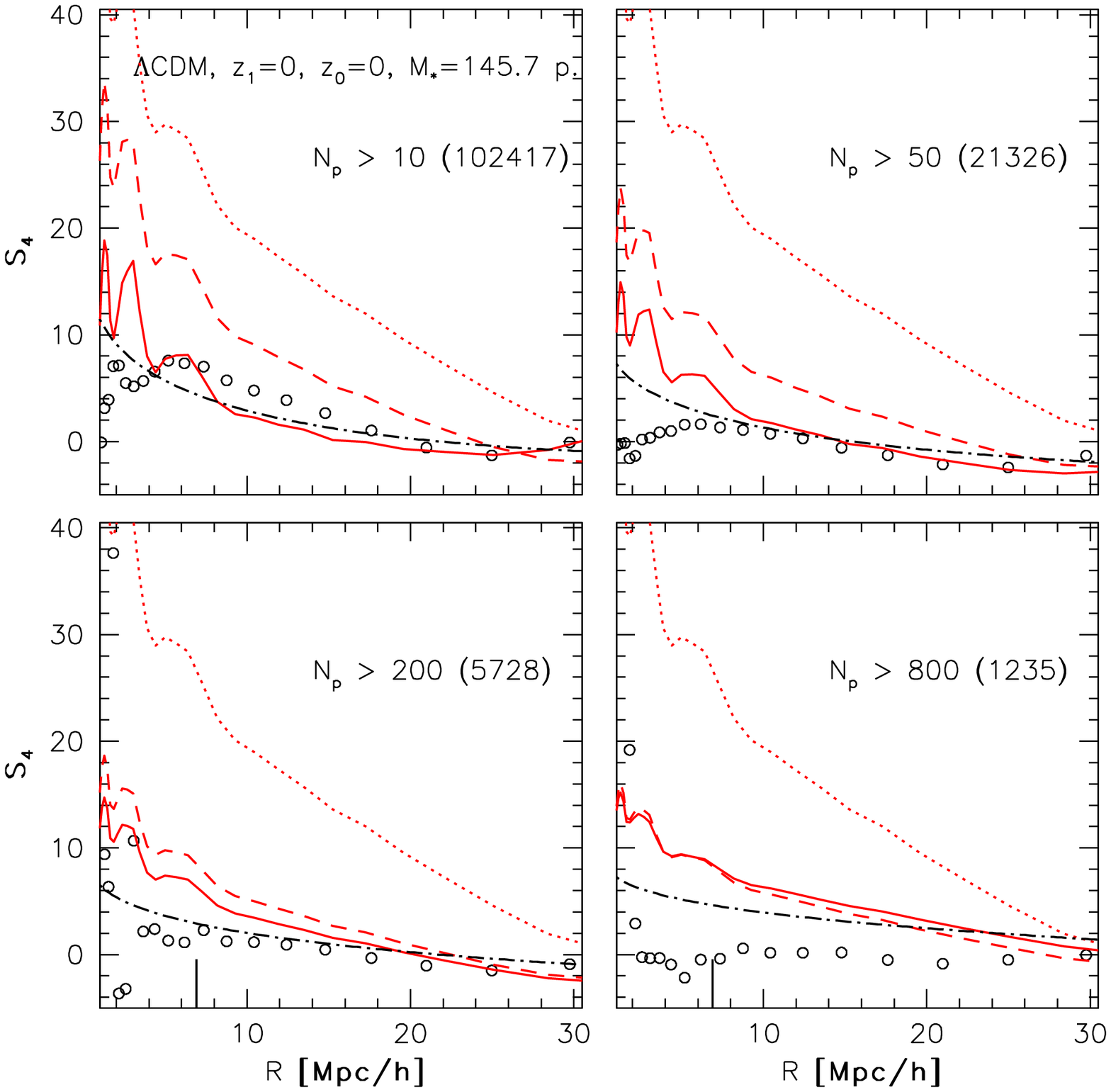}
}
\caption{Skewness $S_3$ and kurtosis $S_4$ of dark haloes with
 different mass ranges obtained from counts-in-cells analysis
 (symbols), from applying the bias model from MJW (solid line) and its
 SMT extension (dashed-line). The moments for the mass distribution are
 shown by the dotted line and the moments for the haloes obtained
 using the moments for the mass from the perturbation theory are shown
 as a dot-long dashed line. The thick ticks on the horizontal axis
 show the scales  where $\bar{\xi}_2 (R) = 1$. Results are shown for
 the VIRGO \lcdm simulations. The haloes have been identified and
 analyzed at the times written in the upper-left boxes. The value of
 $M_{*}$ is also written for more information. Each box corresponds to
 a different range of masses of haloes, as appearing in the
 labels. The quantities between parenthesis correspond to the number
 of haloes in each sample.} 
\label{virgo_sq00}
\end{figure*}

\begin{figure*}
\centerline{
\includegraphics*[bb= 35 160 570 700, width=0.51\textwidth]
{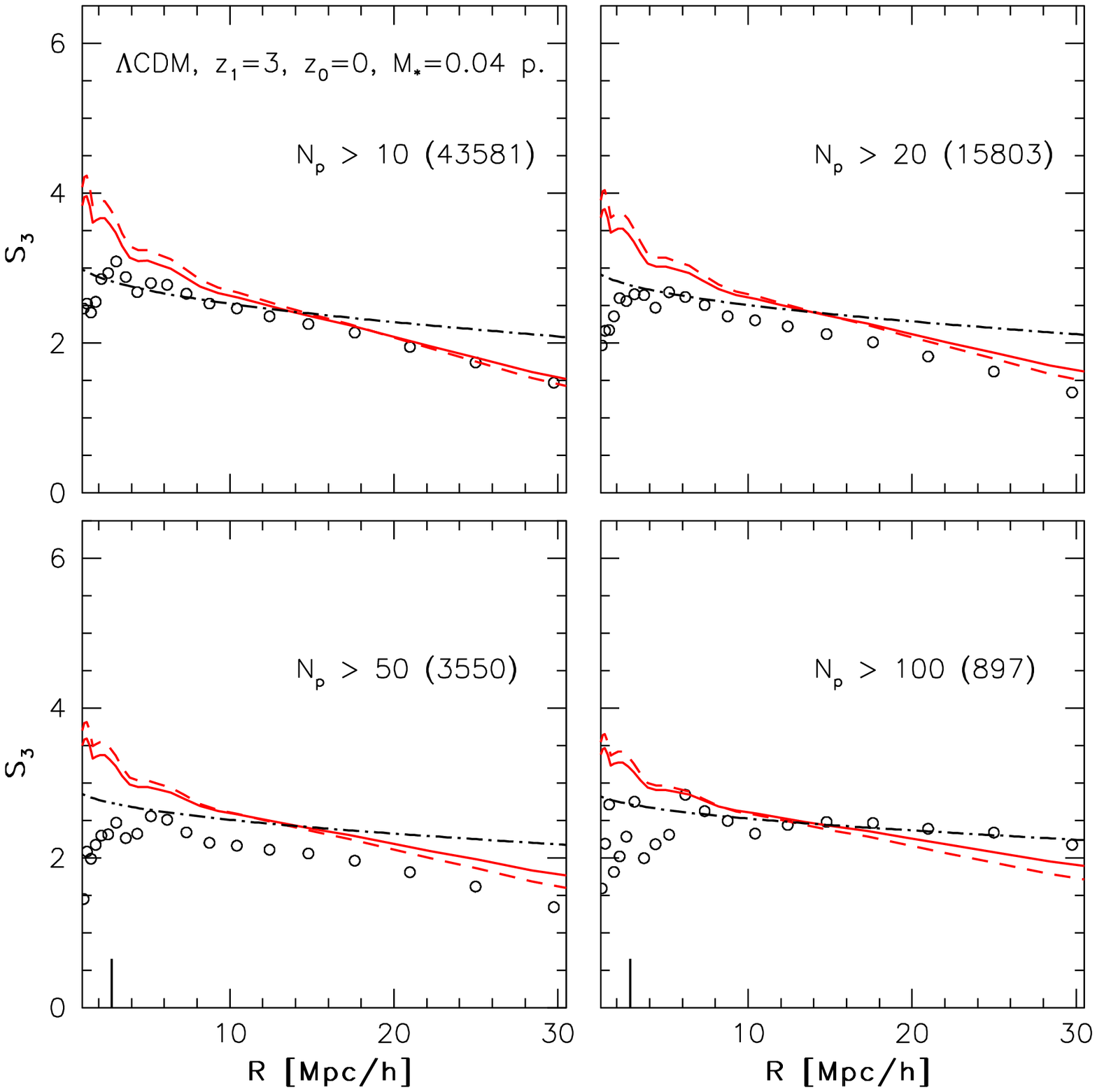}
\includegraphics*[bb= 35 160 570 700, width=0.51\textwidth]
{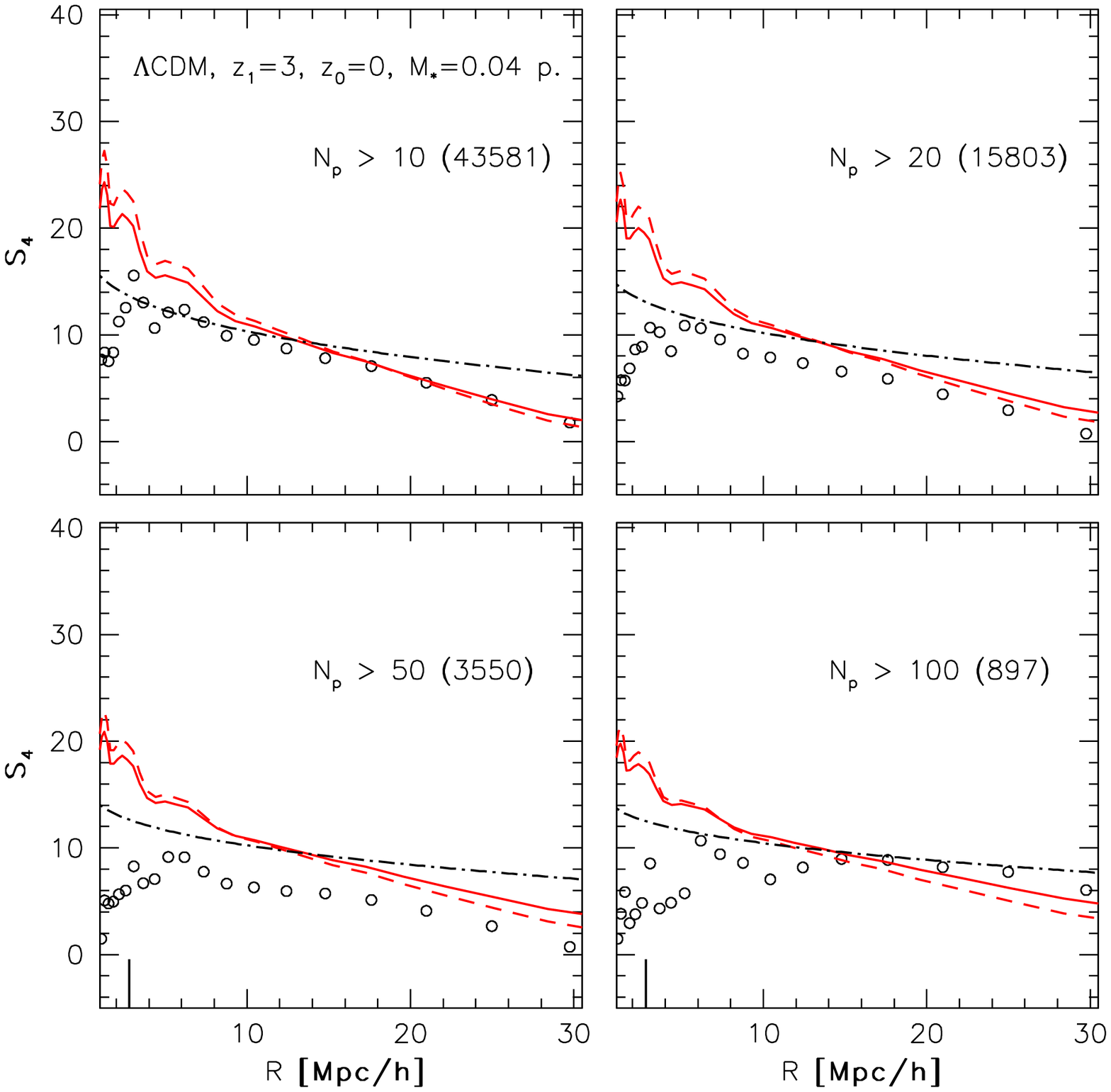}
}
\caption{Skewness $S_3$ and kurtosis $S_4$ for haloes in the VIRGO
\lcdm simulations for haloes identified at $z=3$ and analyzed at
present time. The lines and symbols correspondence is the same as in Figure
\ref{virgo_sq00} as well as the notation.} 
\label{virgo_sq30}
\end{figure*}

In Figures \ref{virgo_sq00} and \ref{virgo_sq30} we show the
third-  and fourth- order moments from the VIRGO \lcdm simulation,
together with model predictions. 
For haloes identified and analyzed at a given epoch (Figure
\ref{virgo_sq00}) both models work remarkably good for scales larger 
than $r_0$ [$\bar{\xi}_2 = 1$], which are the validity scales of
the models, and in particular for haloes with intermediate
masses. Both models are less accurate for very low mass haloes and for very
massive haloes.
On the other hand, in the case of present epoch descendants of haloes
identified at an earlier epoch, both models work remarkably good in
the validity scales of the models, as shown in Figure \ref{virgo_sq30}.

In these two figures we also plot the prediction of the
MJW model with $S_3$ and $S_4$ given by perturbation theory
\citep[see][]{Bernardeau:94}. The fact that this prediction 
also matches the simulations results suggests that the moments 
obtained from the VIRGO simulations are not affected 
significantly by the finite-volume effect and that the
MJW model is a good approximation to the higher-order moments for 
haloes that are not much smaller than $M_\star$ 
[defined by $\sigma(M_\star)=1.68$].

\begin{figure*}
\centerline{
\includegraphics*[bb= 35 160 570 700, width=0.51\textwidth]
{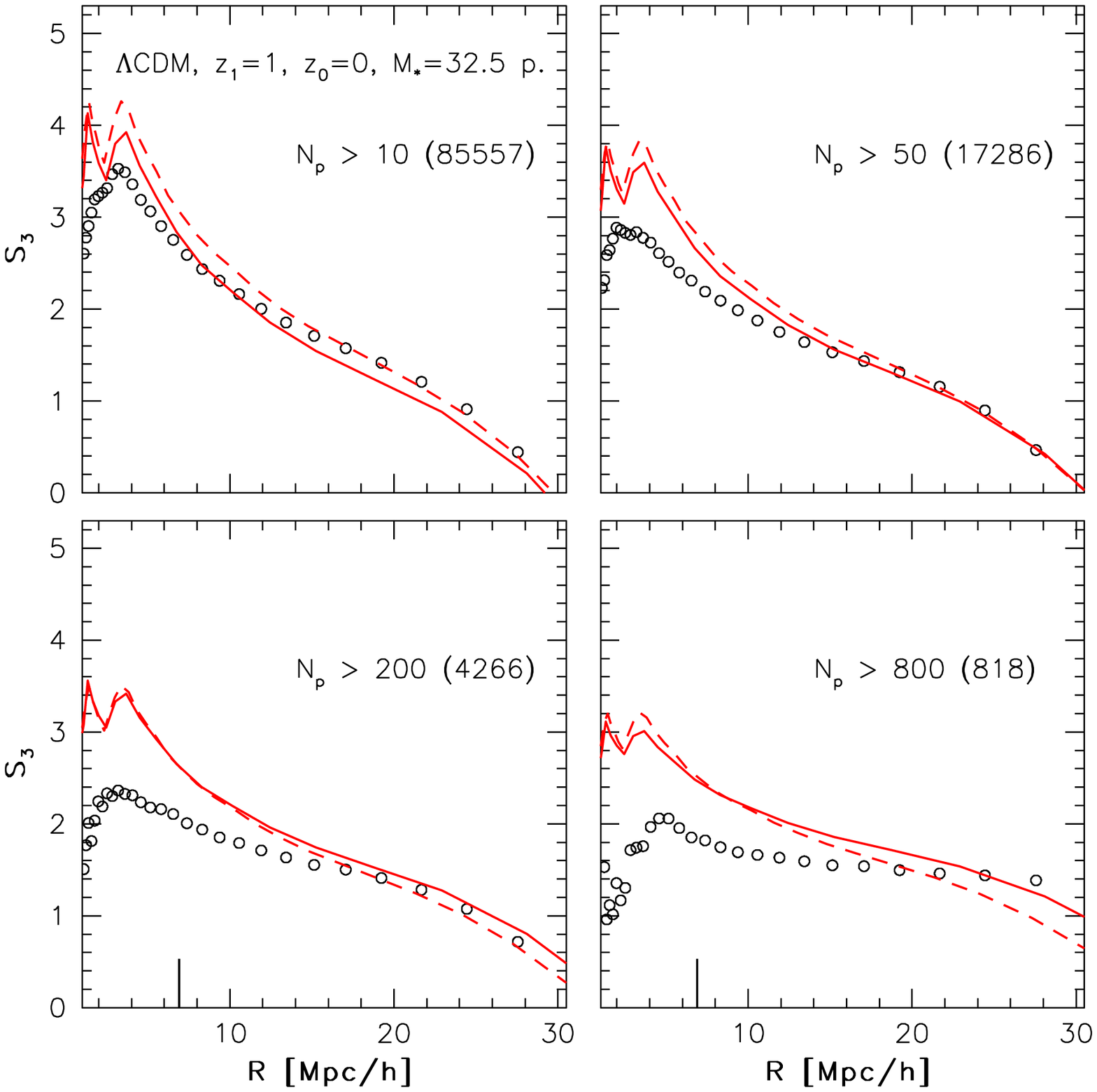}
\includegraphics*[bb= 35 160 570 700, width=0.51\textwidth]
{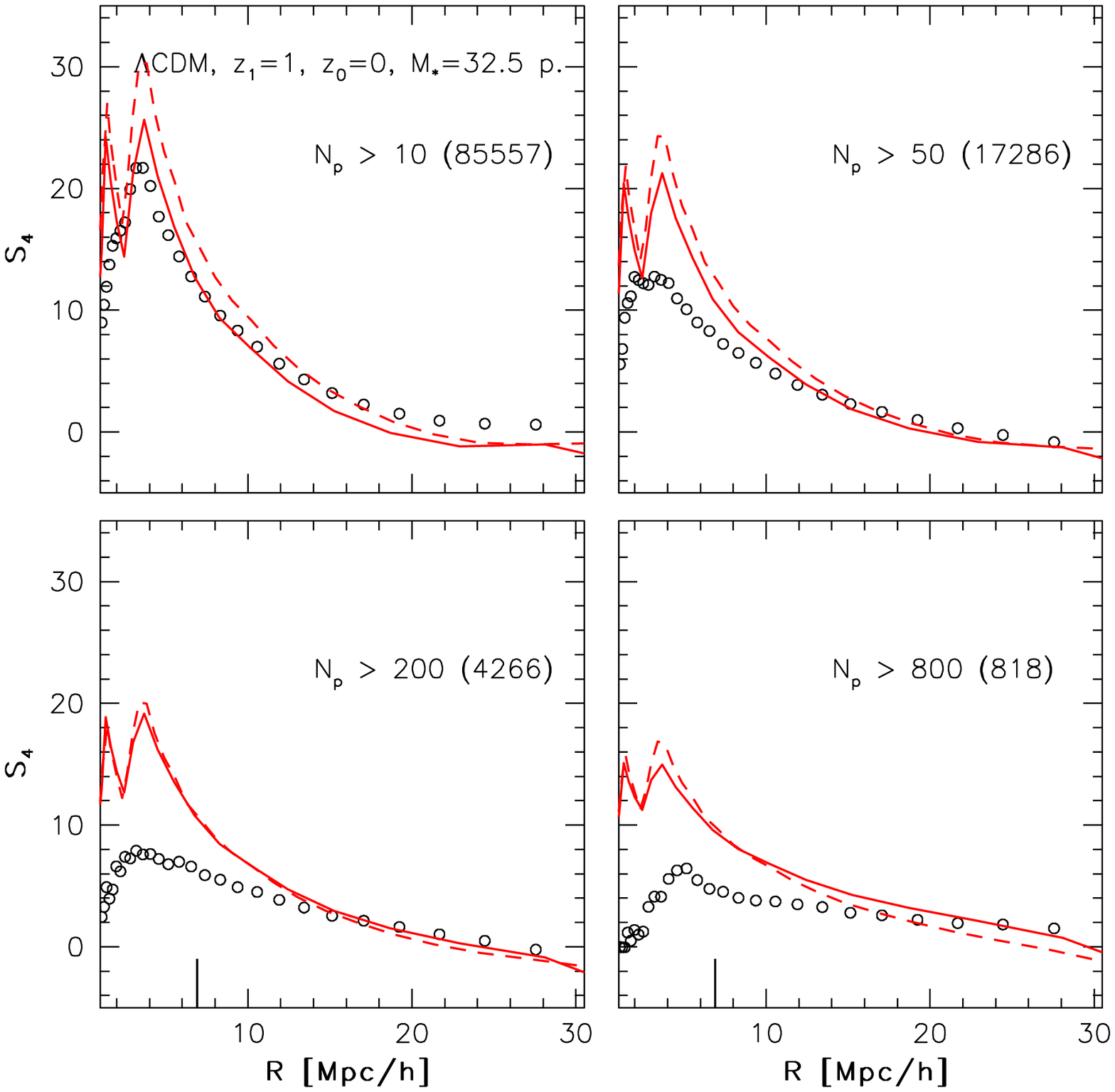}
}
\caption{Skewness $S_3$ and Kurtosis $S_4$ of dark haloes with
 different mass ranges obtained from counts-in-cells analysis (symbols)
 and from applying the bias model from MJW (solid line) and its
 SMT extension (dashed-line). The thick ticks on the horizontal axis
 show the scales  where $\bar{\xi}_2 (R) = 1$. Results  are shown for
 the GIF \lcdm model and for haloes identified at $z=1$ and analyzed at the 
 present time.}  
\label{sq_z1z0}
\end{figure*}

\begin{figure*}
\centerline{
\includegraphics*[bb= 35 160 570 700, width=0.51\textwidth]
{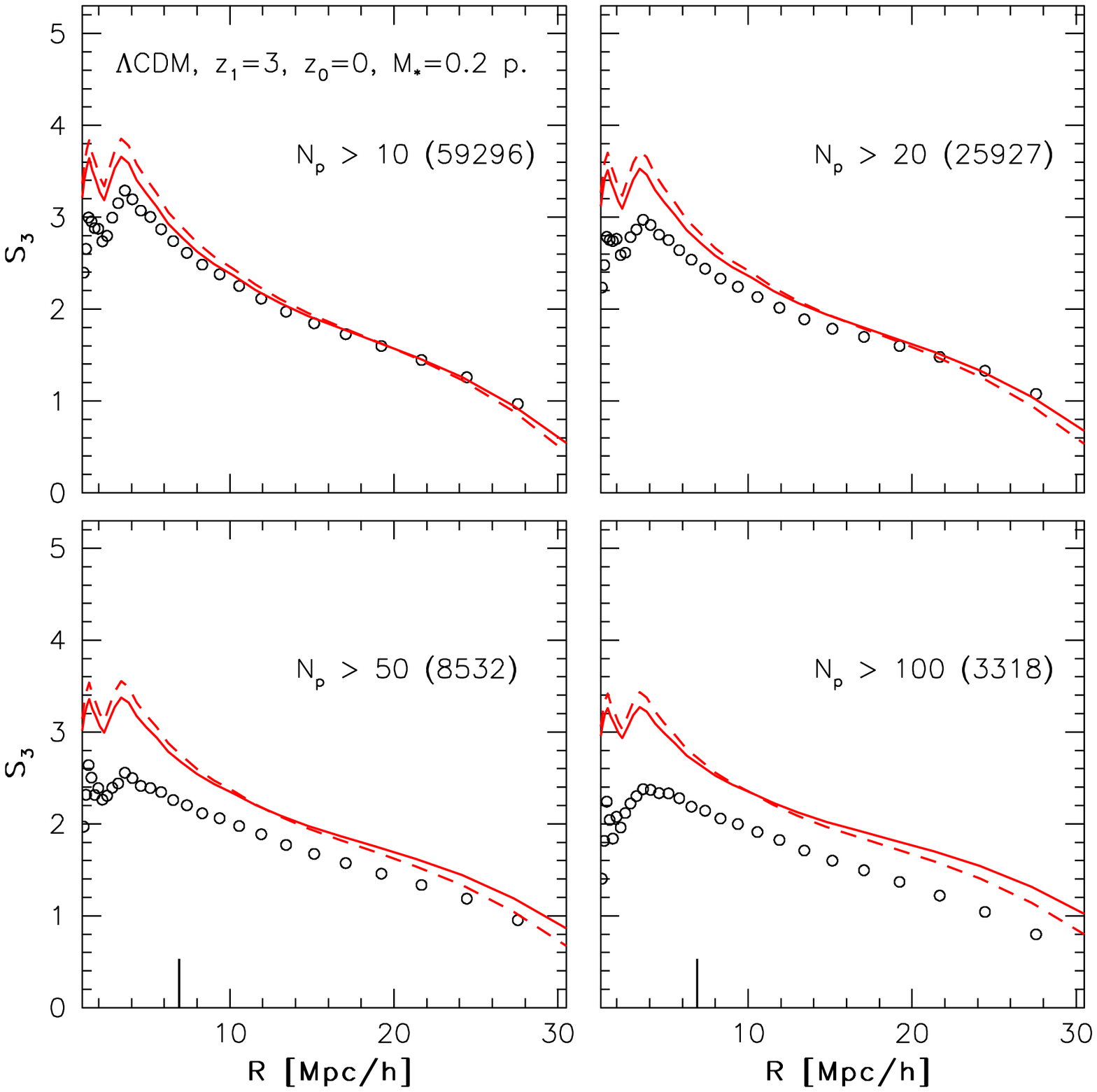}
\includegraphics*[bb= 35 160 570 700, width=0.51\textwidth]
{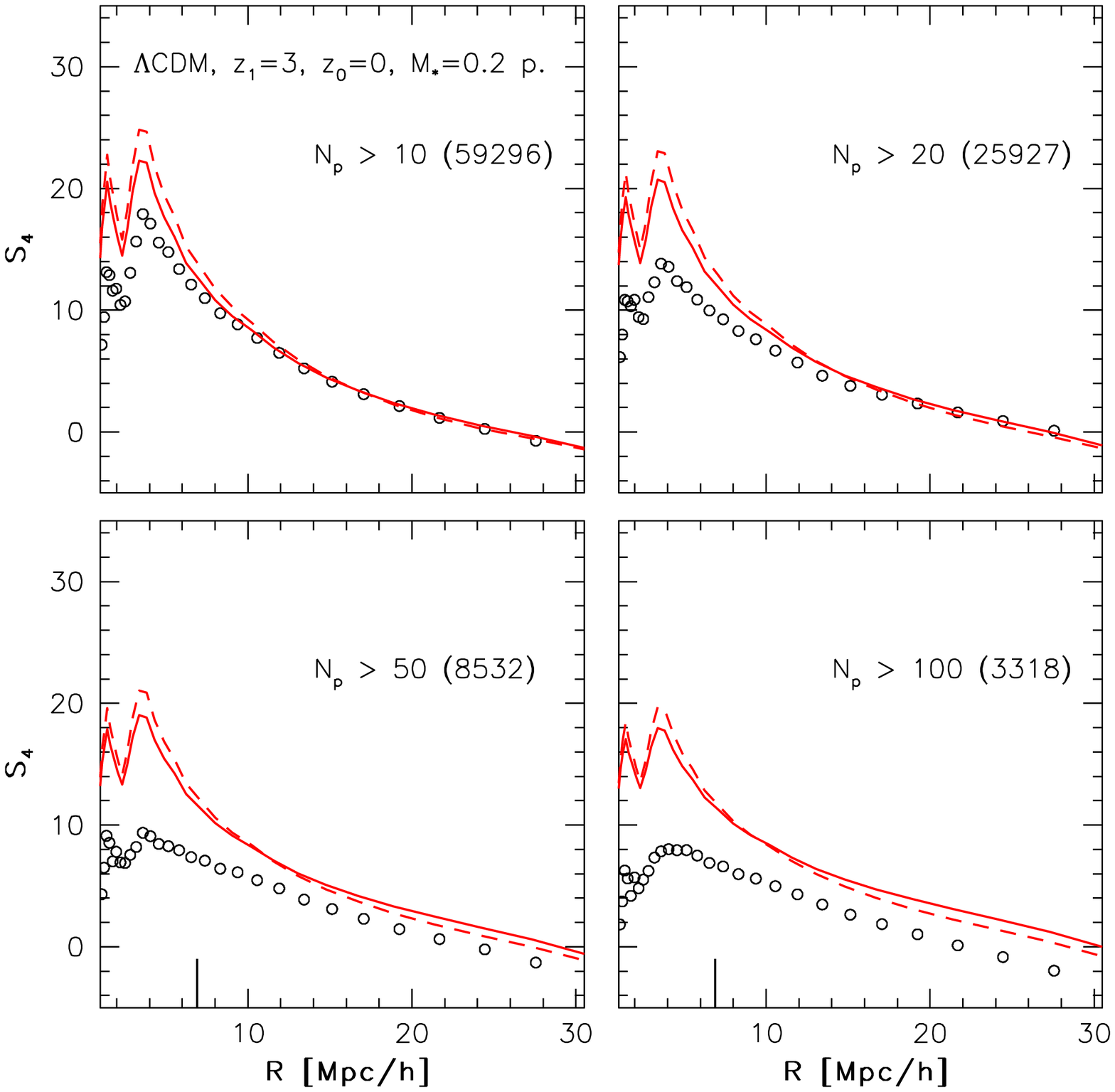}
}
\caption{Skewness $S_3$ and Kurtosis $S_4$ of dark haloes with
 different mass ranges obtained from counts-in-cells analysis (symbols)
 and from applying the bias model from MJW (solid line) and its
 SMT extension (dashed-line).The thick ticks on the horizontal axis
 show the scales  where $\bar{\xi}_2 (R) = 1$. Results 
 are shown for the GIF \lcdm model and for haloes identified at $z=3$ and
 analyzed at the present time.} 
\label{sq_z3z0}
\end{figure*}

With their high mass resolutions, GIF simulations allow us
to test the theoretical models for haloes with mass
$M\ll M_\star$. Since the GIF simulations have relatively 
small simulation boxes, the higher-order moments are expected 
to be affected significantly by the finite-volume effect
\citep[][]{ColombiBS:94}. However, this effect in each simulation is
expected to be similar for both the mass distribution and the halo
distribution. 
Thus, to test the bias model given in (\ref{s3_h}) and (\ref{s4_h})
by a numerical simulation we should use the value of $S_3$ and $S_4$ 
obtained directly from mass distribution in the simulations, because it is the
simulated power spectrum (not the theoretical one)
that is responsible for the clustering in the simulations.
Figures \ref{sq_z1z0} and \ref{sq_z3z0} show the results obtained
for the GIF simulations, along with the predictions from the models,
for present epoch descendants of haloes identified at earlier
epochs. It can be noticed that the predictions from both models, MJW
and SMT, are in good agreement with the simulations results.

As we want to test and compare the performance of the models for low
mass haloes at a given redshift, we have estimated the moments of
counts--in--cells, as well as the moments predicted by the models, for
several populations of sub--$M_\star$ haloes. It is well known that
many of the low--mass haloes identified using a simple
friends--of--friends halo finding algorithm have positive total energy
(i.e., are unbound groups) and that these unbound haloes are expected
to have a significant effect on higher--order statistics
\citep[][]{Benson:01}. For this reason we have revised our GIF halo
catalogues for possible unbound haloes. The total energy for each
halo in the catalog is computed, if it is positive the least bound
particle in the group is removed and the total energy is computed
again. This process is repeated for each halo until the total energy
becomes negative or until there are remaining less than 10 particles
in the group. If the last condition is reached, the halo is removed
from the catalog. We found that $\sim 10\%$ of the haloes in the
original catalog  have to be removed form it, in agreement with
\citet[][]{Benson:01}.

\begin{figure*}
\centerline{
\includegraphics*[bb=32 170 570 705, width=\textwidth]
{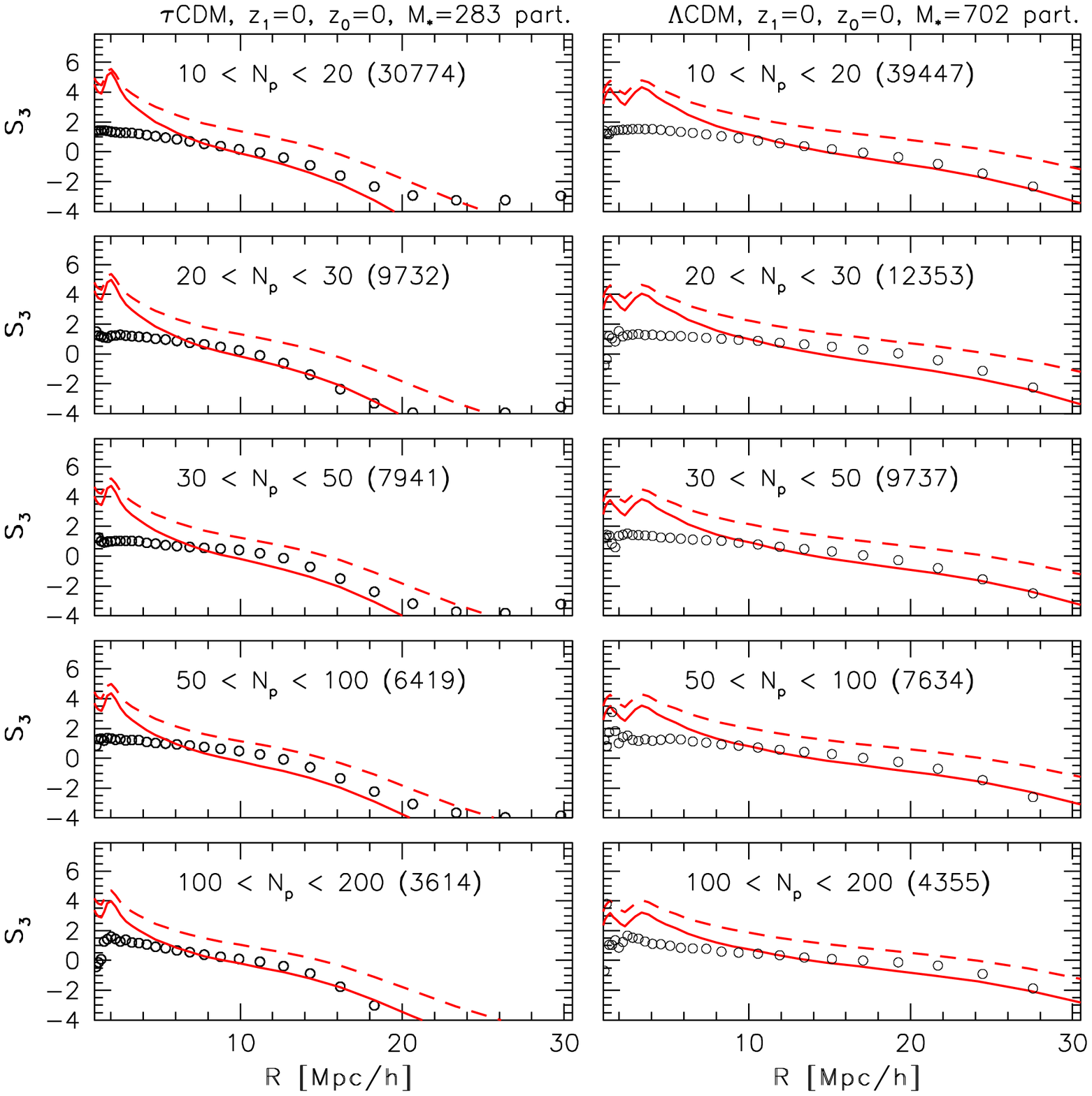}
}
\caption{Skewness $S_3$ obtained from counts-in-cells analysis (symbols)
 and from applying the bias model from MJW (solid line) and its SMT
 extension (dashed-line) of haloes less massive than $M_{*}$. Each row
 in the panel corresponds to a different range of halo masses, as
 indicated in the boxes.} 
\label{s3_binned}
\end{figure*}

\begin{figure*}
\centerline{
\includegraphics*[bb=32 170 570 705, width=\textwidth]
{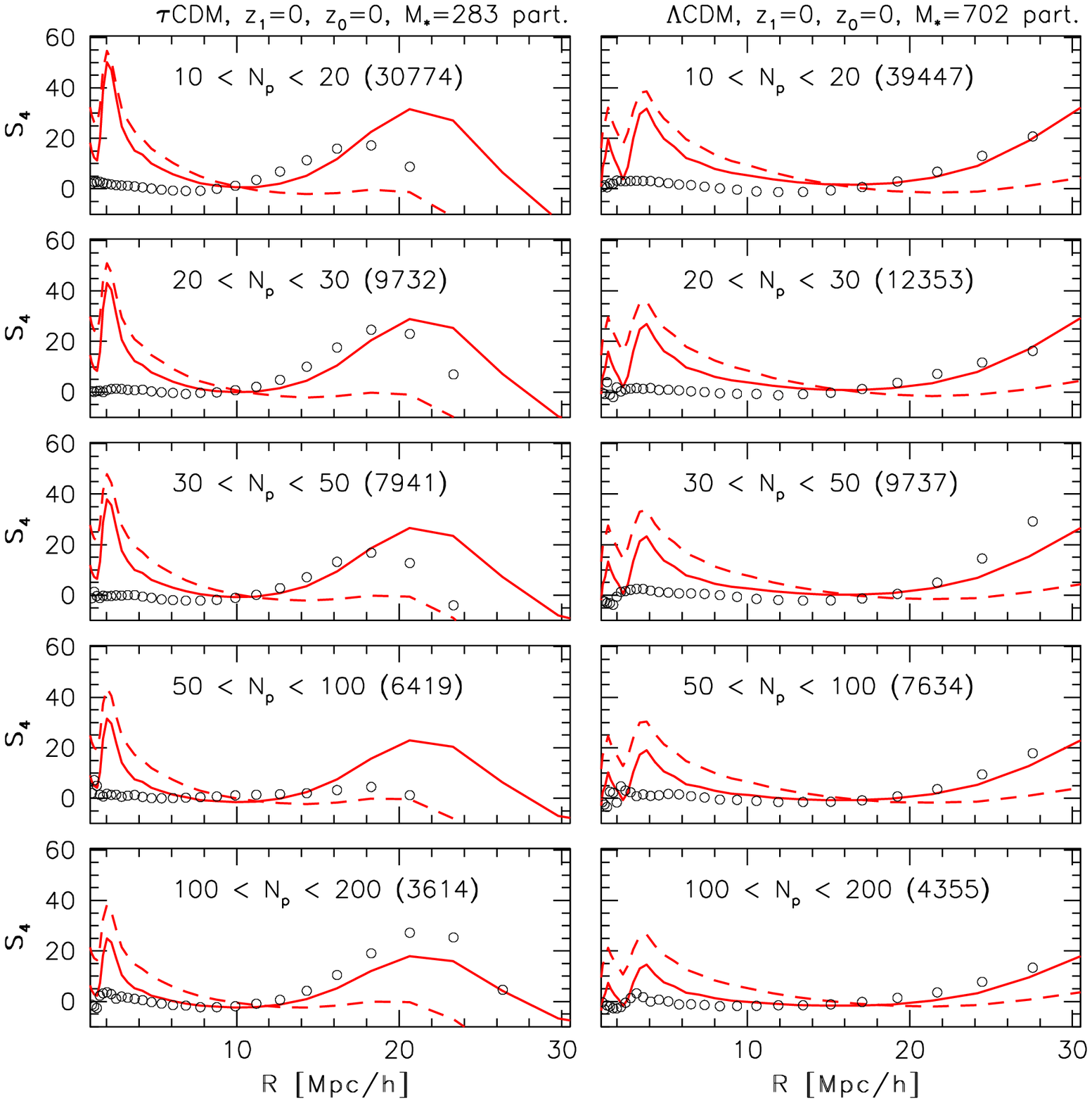}
}
\caption{Kurtosis $S_4$ obtained from counts-in-cells analysis (symbols)
 and from applying the bias model from MJW (solid line) and its SMT
 extension (dashed-line) of haloes less massive than $M_{*}$. Each row
 in the panel corresponds to a different range of halo masses, as
 indicated in the boxes.}
\label{s4_binned}
\end{figure*}

Figures \ref{s3_binned} and \ref{s4_binned} show the results for
present epoch haloes less massive than $M_\star$ in the revised GIF
halo catalogues, along with the predictions from the models. For
haloes with masses much smaller than $M_\star$, the SMT extension
overestimates the skewness and kurtosis, while the MJW model gives a
better fit to the simulations results. Thus, although the SMT
extension improves the models for the mass function and the second
order model of dark haloes, it seems to be unable to improve the
models for the higher--order moments. It is not clear why this
happens, it might be a fluke or it might also mean that the simple
model for the higher--order moments, either spherical or ellipsoidal,
does not work accurately.

% Figure 1 (Virgo 0-0)

%Figure 2 (Virgo 3-0)

%Figure 3
%

% Figure 4

From a comparison between the VIRGO and GIF results,  
it is evident that both the skewness and kurtosis
are strongly affected by the finite-volume effect.
However, if the loss of clustering power due to the finite volume 
is taken into account, the model predictions are in good 
agreement with the numerical results, suggesting that the 
bias relations given by (\ref{s3_h}) and (\ref{s4_h}),
with the coefficients given by the extended Press-Schechter 
formalism, are good approximations to the skewness
and kurtosis of dark haloes in the quasi-linear regime.    

\begin{figure}
\centerline{
\includegraphics*[bb=25 160 570 705, width=\columnwidth]{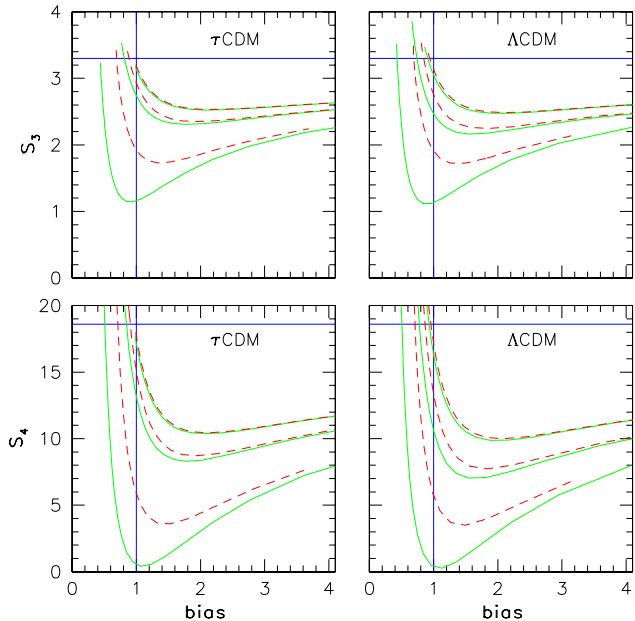}}
\caption{Predictions from the MJW model (solid lines) and its
 SMT extension (dashed-line) for the skewness and kurtosis of haloes at
 a radius $R = 10~h^{-1}~Mpc$ as a function of the linear bias
 parameter $b$. each pair of curves shows the results for a given
 $\delta_1$, where $z_1 \equiv (\delta_1/1.686 -1) = 0., ~1.0,~3.0)$
 from bottom to top.}
\label{sq_10}
\end{figure}

To see more clearly the difference between the MJW model and 
the SMT extension, we show in Figure \ref{sq_10} the
amplitudes of the halo skewness and kurtosis
at a fixed radius ($R=10h^{-1}{\rm Mpc}$) as a function of 
the linear bias parameter $b=b_1$
[see equations (\ref{b1}) and (\ref{b1_st})]. The
curves correspond to the predictions from the models 
for the present-day descendants of haloes at three values
of $z_1$ (3.0, 1.0 and 0.0).  From the figure we see that 
in all cases the values of $S_{j,h}$ are lower than
those for the mass unless $b$ is comparable to or smaller than 1.
This result was obtained in MJW based on the spherical model. 
We see that this is also true even if the SMT extension is used, 
although the amplitudes of $S_{j,h}$ given by the elliptical
model are higher than those given by the spherical model
for a given  $b$. These features in $S_{3,h}$ and $S_{4,h}$ 
have been used in MJW to constrain the bias parameter $b$ for 
galaxies.

% Figure 5

% Figure 6

% Figure 7

% Figure 8
\begin{figure}
\includegraphics*[bb=20 150 570 705, width=\columnwidth]{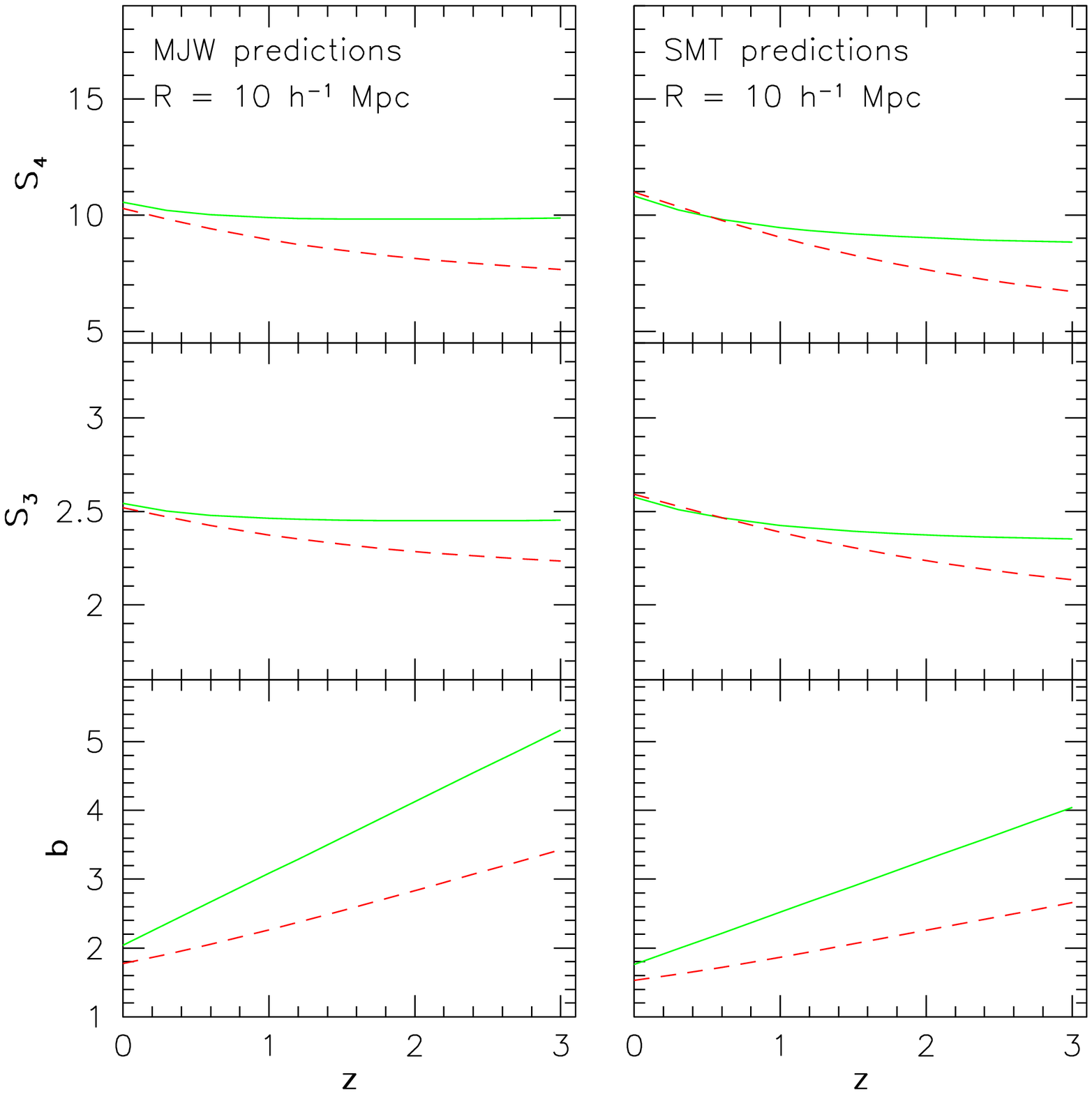}
\caption{Skewness and kurtosis at $R = 10 h^{-1}~Mpc$  for the LBGs at
$z=3$ and their descendants at later epochs. The curves correspond to
the \lcdm model (dashed lines) and the \tcdm model (solid lines). The
left panel shows the predictions from the MJW model and the right
panel shows the predictions from the SMT extension.} 
\label{lbgs}
\end{figure}

\section{Discussion}

From the results shown in Figure \ref{sq_10} we see that for present
time descendants of haloes already formed at a given redshift ($z >
0$), the values of the skewness and kurtosis depend only weakly on the
object mass if the bias parameter $b$ is larger than or near to one . On
the other hand, in the same range of $b$ it is clear that the high order
moments depend on the identification redshift, which is associated
with the redshift of formation of the objects, with the
corresponding values increasing as the formation redshift
increases. Therefore the values of the skewness and kurtosis of old objects,
like elliptical galaxies, are expected to be higher than the corresponding
moments of more recently formed objects, such as spiral galaxies. This
feature can be useful in studying different galaxy populations.

We have used the models to analyze the predicted values of the high
order moments for high redshift objects, like the Lyman Break Galaxies
(LBGs), which are commonly assumed to form in the center of the most
massive haloes at redshift $\sim 3$ \citep[][]{MoF:96, Adelberger:98,
JingS:98, MoMW:99}. Under this assumption and supposing that only a
negligible fraction of those haloes host a secondary observable galaxy
the observed LBGs correspond to the most massive haloes at  
$z \sim 3$. We have estimated the predicted values 
for the skewness and kurtosis at a fixed scale $R = 10~h^{-1}~Mpc$ of
the LBGs ($z=3$) and their descendants at a given redshift $z$. We
chose this value of $R$ because the mass density in the universe is
still in the quasi-linear regime and  the high order moments of galaxy
distributions are more difficult to measure on much larger scales.  

For our estimates we have used the coefficients as given by equations
(\ref{b1})-(\ref{b3}), where the $S_q$ (q = 3,4) for the mass
distribution are obtained from linear perturbation theory
\citep[][]{Bernardeau:94}, and the weighted average needed to get the
effective $b_k$'s is derived either by means of the mass function from the
Press-Schechter formalism, for the predictions from the MJW model, or
by means of the mass function predicted by the SMT model, for the
predictions from the SMT model . The main parameter for the estimation
of the $b_k$'s for the LBGs corresponds to the observed abundance of
LBGs, namely the number density given by
\citep[][]{Adelberger:98}. This number is $N_{lbg} \approx 8 \times
10^{-3} h^3 Mpc ^{-3}$ at $z \sim 3$ for an Einstein-de Sitter
universe, and is similar to the present abundance of $L_*$
galaxies. The corresponding number for the \lcdm universe is estimated
by multiplying this number by the comoving volume per unit redshift at
$z \sim 3$ for an Einstein-de Sitter universe divided by the
corresponding value for the \lcdm universe. 

In Figure \ref{lbgs} we show the values of the skewness, the kurtosis
and the linear bias at $R = 10~h^{-1}~Mpc$ of the LBGs, as a function of
the redshift, in the \lcdm and \tcdm models. From the
curves we see that, although the linear bias parameter is quite different in
both CDM models, the values obtained for the moments are quite similar
and so these statistics do not provide stringent constraints on
cosmological parameters. Note that because LBGs are highly biased
relative to the mass \citep[][]{Adelberger:98}, the skewness and
kurtosis parameters of their haloes and descendants are significantly
lower than those for the mass. Thus, observations on these quantities
may give additional evidence that these objects are highly biased. 

Finally in Figure (9) we show the higher-order moments from the
spatial distribution of model galaxies in the GIF \lcdm simulations at
redshifts ($z=0,\ 1$ and $3$), along with the corresponding
quantities of the distribution of haloes hosting the galaxies and with
the same quantities of the distribution of haloes weighted by the
number of galaxies  hosted by each halo. The
catalogues are limited to model galaxies with stellar masses greater
than $\sim 2 \times 10^{10}~ h^{-1}\ M_\odot$ (i.e., haloes more
massive than $10^{12}~ h^{-1}\ M_\odot$). For further details
about these catalogues and the galaxy formation models used in their
construction see \citet[][]{Kauffmann:99}.

From the plots it can be seen that the skewness and kurtosis
of model galaxies at redshift 1 and 3 is similar to the
corresponding quantities of the dark matter halo population hosting
the model galaxies ($M_h > 10^{12}\ h^{-1}\ M_\odot$). The difference
between the moments of the model galaxies and dark matter haloes are
still appreciable since there are haloes hosting more than one galaxy
and this have strong effects on clustering quantities like the
higher--order moments. Nevertheless, the similarities shown in the
plots suggest that, indeed, one may assume that the distribution of
galaxies may be approximated by the distribution of dark matter
haloes. Indeed from the plot it can be seen that the moments of the
distribution of model galaxies are very similar to the moments of the 
distribution of dark matter haloes weighted by the number of galaxies 
hosted by each halo. At small scales the moments of the weighted halo 
distribution are larger than the moments of the model galaxies 
distribution because we neglect any structure information in a 
halo containing several model galaxies. 

\noindent Thus, one might be able to
describe the  higher-order moments of galaxies by combining the
theoretical models for the higher-order moments of dark matter haloes
with models for the number, stellar mass and position of individual
galaxies that can be hosted by a single dark matter halo with mass $M$
at a given epoch. Furthermore, many of the semi--analytic
model galaxies at redshift 3 would be included in current LBGs samples
and thus are close to a subset of the LBGs population, which is highly
biased respect to the mass. From the figure we see that $S_3 \sim 2$
and $S_4 \sim 5$ at $R=10$ Mpc/h, which is nearly consistent with the
model predictions shown in Figure \ref{lbgs}.

% Figure 9
\begin{figure}
\centerline{
\includegraphics*[width=\columnwidth]{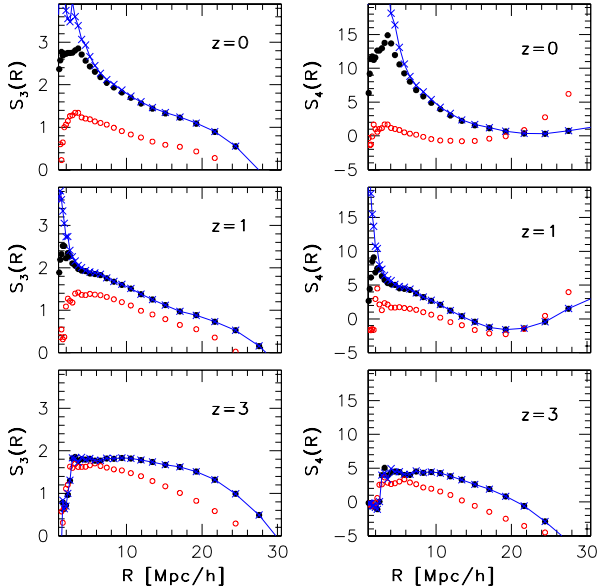}}
\caption{Skewness $S_3$ and kurtosis $S_4$ of the spatial distribution
of galaxies obtained from the GIF \lcdm simulations using
semi--analytical models of galaxy formation (filled circles). $S_3$
and  $S_4$ of the distribution of haloes hosting the galaxies is also
plotted (open circles). For comparison, the respective quantities of
the distribution of haloes weighted by the number of galaxies they
host is included (crosses connected through a line).}  
\label{fig:galaxies}
\end{figure}

\section{Summary} 
We have tested the MJW model and its SMT extension using two sets of
high-resolution N-body simulations with different simulation boxes and
mass resolution. From the set with the very large simulation box, which
allows us to control the finite volume effect, we found that the
models work remarkably good for CDM universes. The good performance of
the models when the moments from the mass distribution are estimated
using the linear perturbation theory, shows that the moments from this
set (VIRGO) are practically unaffected by the finite volume
effect. The other set of simulations, having much higher mass
resolution, has been used to test the model for low-mass haloes,
showing that the model based on spherical collapse (MJW) works  better
than the model based on ellipsoidal collapse (SMT) in describing the
higher--order moments of haloes less massive than $M^*$. For massive
haloes both models work remarkably good.

We use the theoretical models to predict the higher-order moments at a
fixed scale of the Lyman break galaxies observed at $z=3$ and their
descendants at lower redshifts. We found that, although the linear bias
parameter $b$ depends strongly on the cosmology adopted, the values of
the higher-order moments are practically the same in both CDM models and
therefore the higher-order moments from the spatial distribution
of these objects cannot be used to constrain cosmological parameters.

\section*{Acknowledgments}

%We are grateful to Guinevere Kauffmann for a careful reading of the
%manuscript.  
We thank the GIF group and the VIRGO consortium for the
public release of their N-body simulations data 
({\tt www.mpa-garching.mpg.de/Virgo/data\_download.html}).   
R. Casas-Miranda acknowledges financial support from the ``Francisco
Jos\'e de Caldas Institute for the Development of Science and
Technology (COLCIENCIAS)'' under its scholarships program. We thank
our referee for useful comments and suggestions.

\bibliographystyle{mn2e}
\bibliography{Mycitations}

\end{document}